\documentclass[aps,prd,twocolumn,superscriptaddress,secnumarabic]{revtex4-2}

\usepackage[english]{babel}
\usepackage{longtable}
\usepackage{bbm}
\usepackage{enumitem}
\usepackage{graphicx}
\usepackage{xcolor}
\usepackage{wrapfig}
\usepackage{soul}
\usepackage{physics}
\usepackage{hyperref}
\usepackage{multirow}

\definecolor{darkgoldenrod}{rgb}{0.72, 0.53, 0.04}

\usepackage[normalem]{ulem}

\newcommand*{\mcfacts}{\texttt{McFACTS}}
\newcommand*{\cosmic}{\texttt{COSMIC}}

\newcommand*{\OldIMF}{\texttt{OldIMF}}
\newcommand*{\BumpInjection}{\texttt{BumpInjection}}
\newcommand*{\RomExtended}{\texttt{RomExtended}}
\newcommand*{\CosmicSegregated}{\texttt{COSMICSegregated}}

\newcommand*{\pprob}{\mathrm{P}}
\newcommand*{\gwL}{\mathcal{L}}

\newcommand*{\BinaryParameters}{\vec{\lambda}}
\newcommand*{\AllBinaryParameters}{\{\vec{\lambda}\}}
\newcommand*{\FormationParameters}{\Lambda}
\newcommand*{\gwdetj}{d_j}

\newcommand*{\peventj}{\pprob (\gwdetj | \FormationParameters)}

\newcommand*{\bayesfactor}{\mathcal{B}}
\newcommand*{\chieff}{\chi_{\mathrm{eff}}}
\newcommand*{\redshift}{z}
\newcommand*{\metallicity}{\mathcal{Z}}

\newcommand*\mc{\mathcal{M}_c}

\newcommand*{\msun}{M_{\odot}}

\newcommand*{\WeightVolumetric}{\mathcal{R}}
\newcommand*{\WeightIntrinsic}{\rho}
\newcommand*{\WeightDetection}{r}

\newcommand*{\citeLVK}{\cite{LIGO-O4-det,LIGO-O4a-det,Virgo-O3-det,KAGRA-O3-det}}

\newcommand*{\citeOSG}{\cite{osg07,osg09}}
\newcommand*{\citeAstropy}{\cite{astropy:2013,astropy:2018,astropy:2022}}

\newcommand{\AffiliationCCRG}{
  Center for Computational Relativity and Gravitation, 
  Rochester Institute of Technology, 
  Rochester, New York 14623, USA 
}
\newcommand{\AffiliationGSFC}{
  Gravitational Astrophysics Laboratory, 
  NASA Goddard Space Flight Center, 
  Greenbelt, MD 20771, USA 
}
\newcommand{\AffiliationCCA}{
  Center for Computational Astrophysics,
  Flatiron Institute,
  New York, NY 10010, USA
}
\newcommand{\AffiliationAMNH}{
  Department of Astrophysics, 
  American Museum of Natural History, 
  New York, NY 10024, USA 
}
\newcommand{\AffiliationBMCC}{
 Department of Science,
 BMCC, City University of New York,
 New York, NY 10007, USA
}
\newcommand{\AffiliationNMSU}{
  New Mexico State University, 
  Department of Astronomy, 
  PO Box 30001 MSC 4500, 
  Las Cruces, NM 88003, USA 
}
\newcommand{\AffiliationCityUniversity}{
  Graduate Center, 
  City University of New York, 
  365 5th Avenue, 
  New York, NY 10016, USA 
}

\begin{document}
\title{
    Prospects for the formation of GW231123 from the AGN channel
}
\author{V. Delfavero}
\email[]{xevra86@gmail.com}
\affiliation{\AffiliationGSFC}

\author{S. Ray}
\affiliation{\AffiliationAMNH}
\affiliation{\AffiliationCityUniversity}

\author{H. E. Cook}
\affiliation{\AffiliationNMSU}

\author{K. Nathaniel}
\affiliation{\AffiliationCCRG}

\author{B. McKernan}
\affiliation{\AffiliationBMCC}
\affiliation{\AffiliationAMNH}
\affiliation{\AffiliationCityUniversity}
\affiliation{\AffiliationCCA}

\author{K. E. S. Ford}
\affiliation{\AffiliationBMCC}
\affiliation{\AffiliationAMNH}
\affiliation{\AffiliationCityUniversity}
\affiliation{\AffiliationCCA}

\author{J. Postiglione}
\affiliation{\AffiliationAMNH}
\affiliation{\AffiliationCityUniversity}

\author{E. McPike}
\affiliation{\AffiliationAMNH}
\affiliation{\AffiliationCityUniversity}

\author{R. O'Shaughnessy}
\affiliation{\AffiliationCCRG}

\date{\today}

\begin{abstract}

The recent binary black hole (BBH) merger GW231123 consisted of the merger
    of two intermediate mass black holes (IMBH) which appear to have large spin
    magnitudes.
Active galactic nuclei (AGN) are very promising environments for IMBH mergers
    and growth due to high escape velocities.
Here we demonstrate how GW231123 can be produced in the AGN channel.
Using the \mcfacts{} code,
    we explore the impact of various choices of
    the black hole (BH) initial mass function (IMF) on predicted mass and spin
    magnitudes of BBH mergers from the AGN dynamical formation channel.
By integrating the likelihood function for GW231123 with
    the detectable BBH population predicted from AGN using \mcfacts{},
    we demonstrate that GW231123 is consistent with 
    a dynamical BBH merger from the AGN channel.
We also postulate that the masses and spin magnitudes of GW231123
    are most consistent with a merger of fourth and third generation
    BHs, for most choices of a segregated BH IMF and AGN lifetime.

\end{abstract}

\maketitle

\section{Introduction}
\label{sec:intro}
The LIGO-Virgo-KAGRA (LVK) observatories
    \citeLVK{} 
    recently announced the detection
    of a binary black hole (BBH) merger (GW231123)
    with total mass $190 \msun \leq M \leq 265 \msun$
    \cite{GW231123}.
This merger is exceptional for two reasons: First, both black hole (BH) masses
    are consistent with intermediate mass ($\geq 100 \msun$),
    making this the first detected intermediate mass black hole (IMBH) merger.
    Second, both BH spins are consistent with very high spin parameters
    ($\chi_{1}=0.9^{+0.1}_{-0.19}, \chi_{2}=0.8^{+0.2}_{-0.51}$).

The combination of IMBH masses (including $m_{2}$ very probably in the upper mass gap)
    and very high spins is a highly unlikely result of isolated binary evolution \cite{Stegmann2025}.
On the other hand, high mass and high spins are inevitable outcomes in dynamical
    formation channels \cite{McK12,Antonini16,gwastro-popsynVclusters-Rodriguez2016,Mahapatra2025}.
Since merging BH receive a kick, the environment required to explain this observation
    probably has a large escape speed ($\rm{v}_{\rm esc} \gtrsim 150 {\rm km/s}$) \cite{GW231123}.
Such large $\rm{v_{\rm esc}}$ can only be reached in very deep gravitational wells,
    such as close to supermassive black holes (SMBH) in galactic nuclei
    ($\rm{v}_{\rm esc} \sim 10^{3}{\rm km/s}(R/{\rm pc})^{1/2} (M_{\mathrm{SMBH}}/10^8 \msun)^{1/2}$).
Random orientations of mergers
    limit merged BH spin from hierarchical mergers in a gas-free dynamical environment 
    to $a \leq 0.8$ \cite{Kritos24}.
Thus coherent in-plane prograde mergers or gas accretion are required to explain
    the spins in GW231123.
Furthermore, in a gas-free dynamical environment (such as globular clusters), merger rates are expected to be
    lower than in AGN \cite{Ford2022MNRAS}.
Thus active galactic nuclei (AGN) are a very promising environment in which
    GW231123-like events could be produced \cite{McK14,Bartos_2017,Stone17,FordMcK25}.

In this work, we investigate the probability of forming a GW231123-like event in the AGN channel
    as a function of BH initial mass function (IMF),
    using the publicly available \mcfacts{}\footnote{\url{https:www.github.com/mcfacts/mcfacts}}
    (Monte carlo For AGN Channel Testing and Simulations) code
    \cite{mcfactsI, mcfactsII,mcfactsIII}.

We examine the possibility of forming GW231123 in the AGN
    channel following the methods of \cite{mcfactsIII},
    sampling merger populations from a synthetic universe populated by AGN,
    using \mcfacts{} to simulate the formation environment.
For our fiducial model, we make the same assumptions as in that work,
    and also explore the impact of various BH IMF on merger populations.

\section{Methods}
\label{sec:methods}
We sample a synthetic universe and predict BBH mergers from the AGN channel
    detectable by the LVK at O3 sensitivity,
    following \cite{mcfactsIII}.

\subsection{\mcfacts{} models used in this work}
\label{sec:models}
\begin{figure*}[t]
\centering
\includegraphics[width=3.375 in]{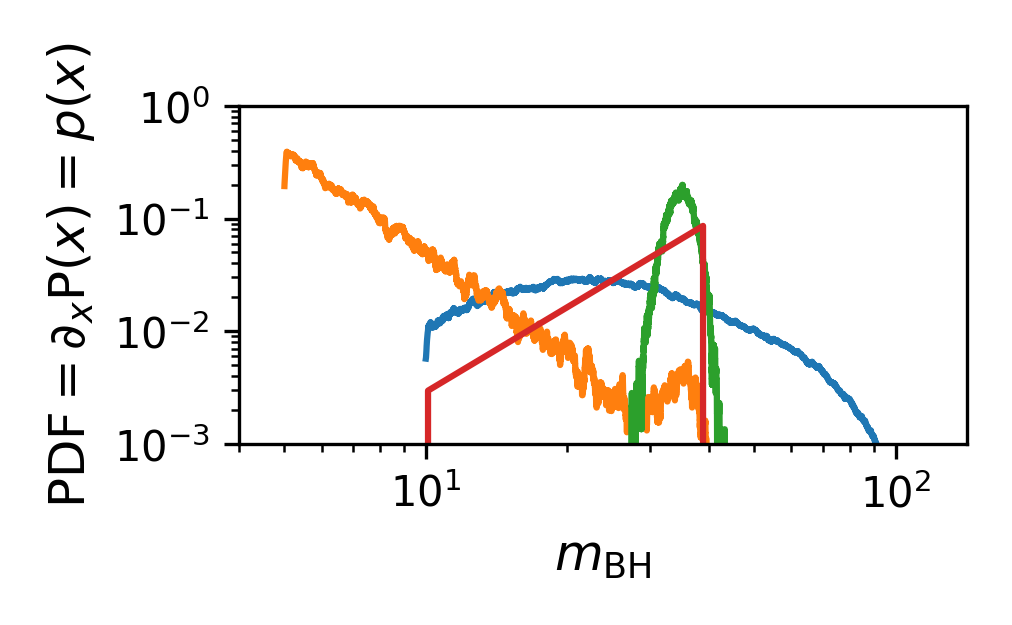}
\includegraphics[width=3.375 in]{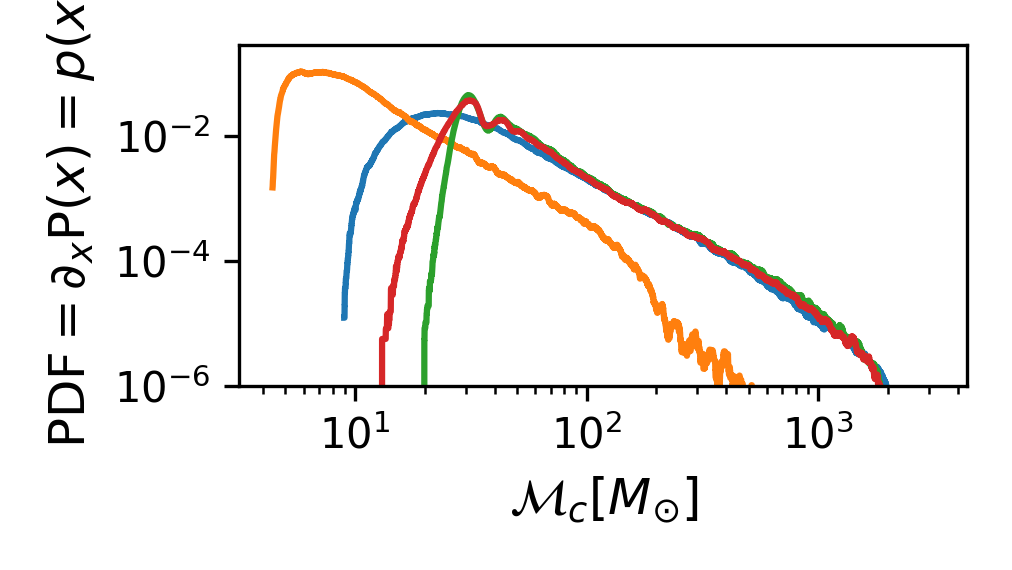}
\caption{\label{fig:mass-spectra}
    \textbf{Black Hole Mass Spectra:}
    (left): 
    The BH IMFs sampled by \mcfacts{} in this work.
    (right):
    The chirp mass spectrum for the BBH merger population (before applying a detection model).
    (both): \OldIMF{} is indicated by orange, \BumpInjection{} by green,
    \RomExtended{} by red, and \CosmicSegregated{} by blue.
}
\end{figure*}

Our AGN disk models follow the scaled Sirko-Goodman (SG) \cite{SG03} models
    presented in \cite{mcfactsIII},
    constructed using the \texttt{pAGN} software package \cite{pAGN}.
We investigate a range of IMFs and disk sizes
    (see Figure \ref{fig:mass-spectra}).
\OldIMF{} corresponds to the IMF used in \cite{mcfactsIII}:
a $m^{-2}$ IMF from $[5,40]\msun$,
    with an additional Gaussian component at $35 \pm 2.3 \msun$ (hereafter \OldIMF{}).
Testing strong mass segregation, \RomExtended{} is
    an IMF with $m^{+2.5}$, spanning $[10,40] \msun$,
    following \cite{Rom2025}.
\BumpInjection{} tests features in the LVK mass spectrum using a Gaussian centered
    on $34 \pm 4 \msun$ \cite{Roy2025}.
Finally, the \CosmicSegregated{} IMF samples BH populations predicted by the 
    \cosmic{} \cite{Breivik2020,COSMIC_code} binary population synthesis code.
\CosmicSegregated{} includes the merged BBH population
    for a synthetic universe consistent with \cite{st_inference_interp}
    and \cite{mcfactsIII}, as well as the BBH components which never merge.
These BH (both the results of BBH mergers and BBH which never merge)
    are weighted consistently with $\rho_k$ as in Eq. 5 of \cite{mcfactsIII},
    with star formation consistent with \cite{st_inference_interp}).
They are further weighted by $m^{1/2}$, consistent with the segregation model
    discussed in \cite{Rom2025}.
Rather than volumetric rates estimated for populations in discrete periods of formation time,
    these ``intrinsic'' rates assume a cosmology model (Planck 2015 \cite{Planck2015})
    to integrate over evolutionary history.

We simulated binaries with \cosmic{}
    in 32 metallicity ($\metallicity$) bins
distributed evenly in logspace between $\metallicity \in [0.0001,0.03]$.
We assume a Kroupa stellar IMF \cite{Kroupa2003},
    and a common envelope efficiency of $\alpha_{\mathrm{CE}} = 1.0$
    with binding energies consistent with \cite{Clayes2014}
    (with ``pessimistic'' Hertzsprung gap assumptions \cite{DominikI}).
We follow the delayed \cite{Fryer2012} supernova remnant mass prescription
    with an electron capture supernova mass range of $[1.4,2.5] \msun$ \cite{Podsi2004}
    and maximum neutron star mass of $3.0 \msun$,
    assuming a ``weak'' pair instability model (applied immediately before core collapse)
    consistent with \cite{Belczynski2020-EvolutionaryRoads}.
Consistent with \cite{st_inference_interp},
    this model draws kicks from a single Maxwellian with
    $\sigma_{\mathrm{kick}}=108.299\mathrm{km/s}$
    (which are not reduced by fallback),
    assumes an efficiency of accretion during Roche-lobe overflow of $0.922$,
    and assumes a reduction in wind mass loss while a hydrogen envelope remains.
This reduction is implemented as a factor of 3, compared to \cite{Vink2001}
    (roughly consistent with \cite{st_inference_interp}, but also with EM observations
    \cite{Cohen2014,GormazMatamala2022,Gayley2022}
    and simulation \cite{Bjorklund2023,Merritt2025}).

We use the \texttt{precession} package
    \cite{precession} to predict the merged
    remnant masses, kicks, and spins for BBH merger products in \mcfacts{}
    in this work.

\begin{figure*}[t]
\centering
\includegraphics[width=3.375 in]{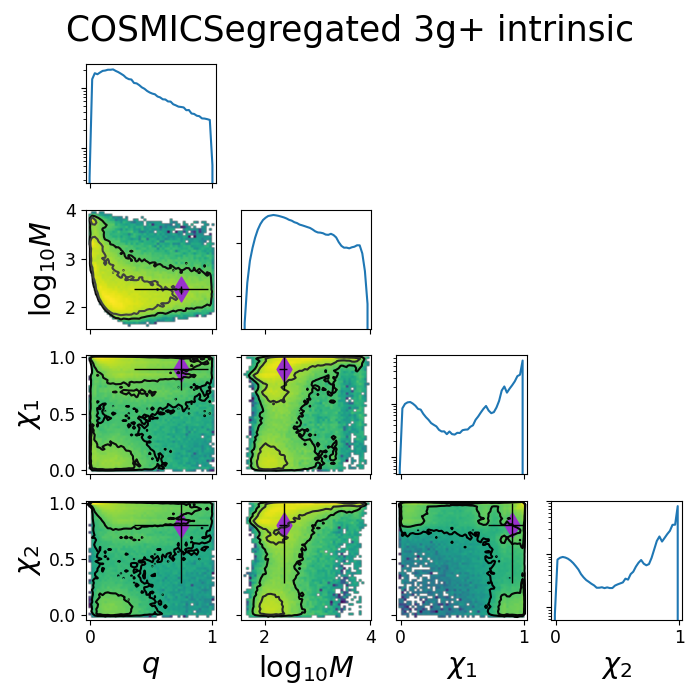}
\includegraphics[width=3.375 in]{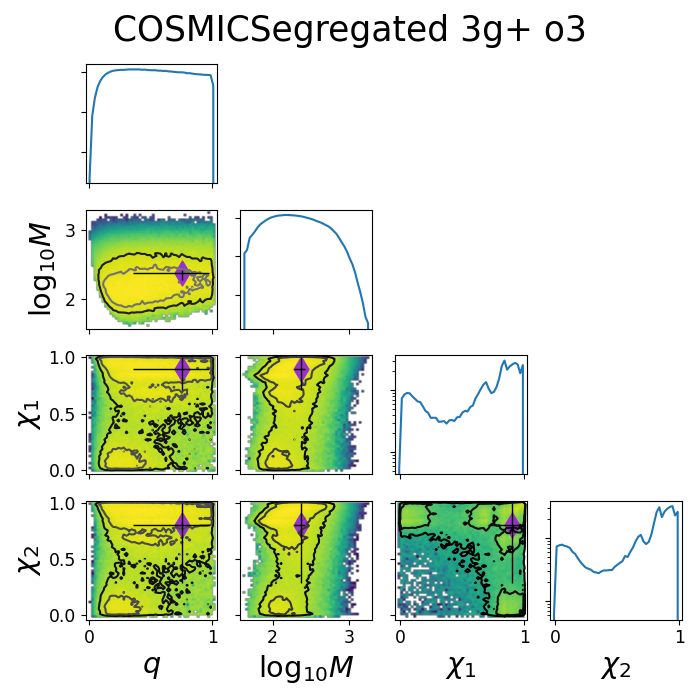}
\caption{\label{fig:corner-plots}
    \textbf{Masses and Spin Magnitudes of AGN BBH}:
    Each of these corner plots illustrates one one- and two-dimensional
        weighted histograms for the mass ratio ($q$),
        total mass ($M$), spin magnitude of the primary ($\chi_1$),
        and spin magnitude of the secondary ($\chi_2$) for the
        hierarchical (third-generation and above) BBH mergers of 
        predicted from \mcfacts{} BBH mergers in a synthetic universe.
    Each sample is weighted by the
        (left) intrinsic weight ($\WeightIntrinsic{}$) /
        (right) detection weight ($\WeightDetection{}$); see Section \ref{sec:postproc}.
    Two-dimensional histograms include $68\%$ and $90\%$ confidence intervals,
        as well as a scattered purple point with expected value and error bars
        consistent with GW231123 \cite{GW231123}.
}
\end{figure*}

\subsection{Sampling BBH in a synthetic universe of AGN}
\label{sec:postproc}
Following \cite{mcfactsIII},
we divide the history of our Universe into 100 $\mathrm{Myr}$ epochs
    and assume an AGN number density (AGND) consistent with the 
    $\log_{10}(L/L_{\odot}) \in [10.5,11]$ luminosity bin from \cite{Lyon2024}
    to estimate the number of AGN per $\mathrm{Mpc}^3$.
We sample the galactic stellar mass function (GSMF) from \cite{Fontana2006}
    and the average metallicity of star formation as a function of redshift from \cite{Madau2017},
    with $0.5$ dex uncertainty.
We use the stellar mass and metallicity of the galaxy to classify early- or late-type
    galaxies by finding the closest match to either curve from Figure 2a in \cite{Peng2015}.
Then, we use a power law to determine the SMBH
    mass \cite{Schramm_2013} and nuclear star cluster (NSC) mass \cite{Neumayer2020}
    of each galaxy sample.

Drawing $10^{4}$ samples from each epoch out to $\redshift=2$,
    we estimate the relative contribution of the batch of 100 \mcfacts{}
    simulated galaxies representing a particular mass bin
    (distributed in 33 stellar mass bins $[10^{9},10^{13}] \msun$).
The volumetric rate within each epoch ($\WeightVolumetric{}_k$) represented by each 
    BBH merger $k$ from each simulation
    is given by Eq. 4 of \cite{mcfactsIII}.
Assuming isotropy and a Planck 2015 \cite{Planck2015} cosmology,
    we integrate the various epochs over comoving volume
    and estimate the intrinsic rate ($\WeightIntrinsic{}_{k}$) represented by each sample
    within our synthetic universe
    (implemented using Astropy \cite{astropy:2013,astropy:2018,astropy:2022}).
The equation for calculating $\WeightIntrinsic{}_{k}$ is given by Eq. 5 of \cite{mcfactsIII}.

To estimate the rate of detections, we estimate an optimal matched-filter signal-to-noise ratio (SNR)
    for a single-detector with O3 sensitivity using the IMRPhenomPv2 \cite{IMRPhenomPv2}
    waveform model implemented in lalsuite \cite{lalsuite},
    with the \texttt{SimNoisePSDaLIGOaLIGO140MpcT1800545} point spread distribution
    \citep{P1200087}.
We estimate a probability detection $p_{\mathrm{det}}(8/\mathrm{SNR}_k)$
    consistent with \cite{DominikIII} for each sample $k$ (for $\mathrm{SNR}>8$),
accounting for isotropicly distributed sky locations and orientations.
We use a volume-time (VT) estimate in our detection model following \cite{T1800427}
    (implemented in \url{https://git.ligo.org/daniel.wysocki/bayesian-parametric-population-models}),
    where the adjustment freezes $m_1$ and $m_2$ at $100 \msun$ to avoid extrapolation.
The rate of detections represented by a single sample $k$ is then given by
    $\WeightDetection{}_k = \WeightIntrinsic{}_k \mathrm{VT}(\BinaryParameters_k) p_{\mathrm{det}}(8/\mathrm{SNR}_k)$,
    where $\BinaryParameters_k$ are the parameters of each sample
    (mass, spin, redshift, etc.).

\subsection{Bayesian event likelihoods}

Following \cite{st_inference_interp},
    we characterize the probability $\peventj$ of a given gravitational wave (GW)
    observation $\gwdetj$ given a particular model characterized by formation assumptions
    $\FormationParameters$.
The likelihood function characterizing a particular GW observation $j$
    can be represented by
    $\gwL_j(\BinaryParameters) = \pprob(\gwdetj|\BinaryParameters, \FormationParameters)$.
Following this, 
\begin{equation}\label{eq:likelihood-abstract}
\peventj = \int\limits_{\AllBinaryParameters}
    \pprob{}(\gwdetj|\BinaryParameters,\FormationParameters)\pprob{}(\BinaryParameters|\FormationParameters)
    \mathrm{d}\BinaryParameters =
    \int\limits_{\AllBinaryParameters}\bar{\WeightDetection}(\BinaryParameters)\gwL(\BinaryParameters)
    \mathrm{d}\BinaryParameters .
\end{equation}
Here, we normalize $\pprob{}(\BinaryParameters|\FormationParameters) = \bar{\WeightDetection}(\BinaryParameters) =
    \WeightDetection(\BinaryParameters) / \int_{\AllBinaryParameters} \WeightDetection(\BinaryParameters) \mathrm{d} \BinaryParameters$.
Thus we find
\begin{equation}\label{eq:likelihood-discrete}
\peventj = \frac{\sum\limits_k \WeightDetection_k \gwL(\BinaryParameters_k)}{
    \sum\limits_k \WeightDetection_k}.
\end{equation}
We use these event likelihoods to estimate the Bayes factor $\bayesfactor(A,B)$
    between two formation channels characterized by assumptions $\FormationParameters_{\mathrm{A}}$
    and $\FormationParameters_{\mathrm{B}}$.

Single event likelihoods have previously been used to infer the evolutionary pathways
    for specific events \cite{Kimball2019,Mahapatra2021,Mould2023,Mahapatra2024},
    including dynamical merger hierarchies \cite{AraujoAlvarez2024}.
Here, we characterize the likelihood function for a GW event by
    fitting a truncated multivariate normal distribution in 
    chirp mass ($\mc = (m_1m_2)^{3/5}/(m_1+m_2)^{1/5}$), symmetric mass ratio ($\eta=m_1m_2/(m_1+m_2)$), and spin magnitudes
    ($\chi_1$ and $\chi_2$)
    to the likelihood provided for each posterior sample
    in \cite{GW231123-zenodo}
    (for the NRSur7dq4 waveform model \cite{NRSur7dq4}).
BBH spin mis-alignment can result from spheroid interactions in an AGN disk that
    has not yet captured the spheroid population \cite{FordMcK25}.
However, due to disagreement between the posterior sample distributions for GW231123 in the 
    $\chieff$ and $\chi_p$ coordinates \cite{GW231123},
    we focus on spin magnitudes in this work.
The Gaussian fit and the corresponding version of the \texttt{GWALK}
    \cite{nal-chieff-paper,nal-methods-paper} 
    package are included in the accompanying data release.

\section{Results}
\label{sec:results}
\begin{figure}[t]
\centering
\includegraphics[width=3.375 in]{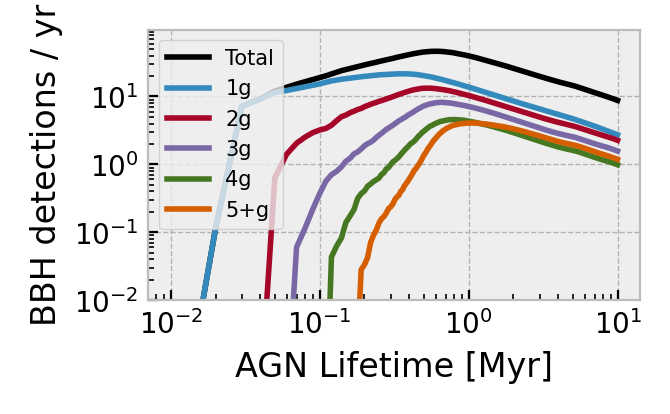}
\includegraphics[width=3.375 in]{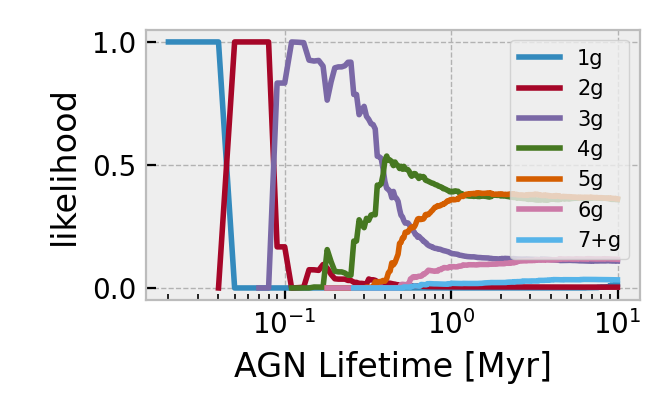}
\caption{\label{fig:lifetime}
    \textbf{Dependence on AGN lifetime:}
    (top): The integrated detection rate as a function of AGN lifetime.
    (bottom): The probability that each generation of merger is responsible
        for GW231123 (assuming it did come from the AGN channel)
        as a function of AGN lifetime.
    The second-generation bump around 0.2 Myr is a result of the (2g,2g) mergers;
        previously, the 2g line is dominated by (2g,1g) mergers.
    Both plots assume the \CosmicSegregated{} BH IMF.
    These rate estimates are estimated with underlying assumptions including that all AGN
        have a scaled Sirko-Goodman \cite{SG03,mcfactsIII} disk and an identical lifetime. 
    Predicted rates are strongly scaled by choices of those assumptions (see Table \ref{tab:rates}),
        and are therefore useful in eliminating or modifying models
        which are constrained by observed rates.
}
\end{figure}

Our results demonstrate that it is possible for a BBH merger with mass and spin properties
    like GW231123 to  form dynamically in the accretion disk of an AGN.
The right-hand panel of Figure \ref{fig:mass-spectra} illustrates the mass spectrum for a synthetic universe
    of AGN disk BBH mergers (simulated using \mcfacts{}) with the BH IMF choices discussed
    in Section \ref{sec:models}, demonstrating coverage in the vicinity of $\mc=100 \msun$
    for all choices of IMF.
However, we do see stronger coverage (by about an order of magnitude) for
    all three IMF choices considering mass segregation (or an injected high-mass feature).

Figure \ref{fig:corner-plots} illustrates the density of mass and spin properties of high-generation 
    BBH mergers (with at least one third-generation (3g) or higher component)
    in the AGN formation channel using the \CosmicSegregated{} BH IMF.
We see that in all the two-dimensional histograms,
    GW231123 (scattered in purple) either overlaps with or has error bars overlapping with
    the 90\% contour line, indicating that GW231123 is entirely consistent with
    AGN-channel predictions without a high event-rate.

Figure \ref{fig:lifetime} shows the rate of BBH merger detections (at O3 sensitivity)
    and the probability of each generation of merger 
    as a function of AGN lifetime (assuming GW231123 did come from the AGN channel).
AGN lifetimes could range up to $100$ Myr,
    however, observational evidence favors shorter lifetimes
    ($0.1-10$ Myr) \cite{Schawinski2015MNRAS,Davies2020MNRAS}
    evidence favors shorter lifetimes (
Because a ``1g BH'' in our analysis of dynamical
    BBH mergers in AGN disks is any sample from the BH IMF,
    this may differ from the total merger generation
    of BHs sampled from the \CosmicSegregated{} IMF,
    some (but not all) of which are isolated binary merger products.
While absolute rates are still sensitive to many of our assumptions
    (such as our disk model; see \cite{mcfactsI,mcfactsII,mcfactsIII}),
    the relative rates of different generations of AGN mergers
    and their dependence on AGN lifetime are valuable
    takeaways from the top panel of Figure \ref{fig:lifetime}.

\begin{table}
\begin{tabular}{|c|cccc|}
\hline
\multirow{2}{*}{IMF} &
\multicolumn{4}{c|}{$\mathrm{detections}/\mathrm{yr}\ [\mathrm{T}_{\mathrm{AGN}}]$} \\
 &
$0.5\,\mathrm{Myr}\ $ &
$1.0\,\mathrm{Myr}\ $ &
$2.0\,\mathrm{Myr}\ $ &
$10.0\,\mathrm{Myr}\ $ \\
\hline
\OldIMF{} & 0.9 & 1.6 & 2.5 & 1.5 \\
\CosmicSegregated{} & 46.3 & 39.8 & 26.0 & 8.7 \\
\RomExtended{} & 82.1 & 60.8 & 38.0 & 11.3 \\
\BumpInjection{} & 90.8 & 65.7 & 40.7 & 11.9 \\
\hline
\end{tabular}
\caption{\label{tab:rates}
\textbf{Detection rates for each IMF:}
The predicted number of detections per year from the AGN channel
    for each choice of BH IMF, at various AGN lifetimes.
We note that these rates depend critically on many assumptions
    (see \cite{mcfactsIII} and Section \ref{sec:discussion} of this work).
}
\end{table}
As indicated by Table \ref{tab:rates} and the top panel of Figure \ref{fig:lifetime},
    we demonstrate that the AGN lifetime and BH IMF has a significant impact
    on the predicted rate of BBH mergers from the AGN channel.

The probabilities estimated in the bottom panel of Figure \ref{fig:lifetime}
    are calculated by evaluating the likelihood integral
    (Eq. \ref{eq:likelihood-discrete})
    at discrete delay time limits between 0 and the maximum simulated AGN lifetime (10 Myr)
    for subpopulations of different generation, and are comparable to Bayes factors 
    between different generations of mergers.
If we hold \textit{prior} expectations that GW231123 is a 4g or higher merger
    in an AGN disk, 
    we conclude based on Figure \ref{fig:lifetime} that for the \CosmicSegregated{} IMF,
    \mcfacts{} predicts that the minimum AGN lifetime
    for the AGN containing GW231123 is $0.2-0.4$ Myr.

\begin{table*}
\begin{tabular}{|c|c|c|c|c|c|c|c|}
\hline
\OldIMF{} & P & \CosmicSegregated{} & P & \RomExtended{} & P & \BumpInjection{} & P \\
\hline
(8g, 7g)    & 0.101 & (4g,3g) & 0.188 & (4g,3g) & 0.504 & (4g,3g) & 0.642 \\
(9g, 8g)    & 0.091 & (5g,4g) & 0.183 & (4g,4g) & 0.260 & (3g,3g) & 0.190 \\
(9g, 9g)    & 0.081 & (4g,4g) & 0.155 & (3g,3g) & 0.125 & (4g,4g) & 0.141 \\
\hline
\end{tabular}
\caption{\label{tab:generation}
\textbf{Probabilities for various hierarchical merger generations:}
    Likelihoods of each combination of merger products for producing a detectable
        BBH merger, given each set of assumptions for a 10 Myr AGN lifetime.
    The top 3 probable modes are shown, with probabilities rounded
to the third decimal place.
    Here, a (1g,1g) merger would be a merger of two BHs sampled
        from the IMF and a (N,M) merger would consist of two BHs
        which are dynamical merger products in the AGN disk,
        from previous mergers having a highest BH generation of $\mathrm{N}-1$ and $\mathrm{M}-1$, respectively.
    The (4g,3g) mode dominates the \RomExtended{} and \BumpInjection{} probablities,
        while the bottom-heavy \OldIMF{} requires higher generation mergers.
}
\end{table*}

Table \ref{tab:generation} further breaks down the probabilities of each combination of BBH generations
    (assuming GW231123 did come from the AGN channel).
For sufficiently long AGN lifetimes, a highly segregated BH IMF predicts that events like GW231123
    strongly favor the (4g,3g) mode.
As the likelihood integration is normalized for the overall population of BBH samples
    from a given \mcfacts{} model, these probabilities implicitly depend on both the properties
    (mass and spin) of sample mergers and the fraction of mergers from each subpopulation.
The \CosmicSegregated{} IMF slightly favors the (4,3) mode for an AGN lifetime of 10 Myr;
    Figure \ref{fig:lifetime} shows how the dominance of different generations
    changes with the assumed AGN lifetime.

\section{Discussion}
\label{sec:discussion}

The LVK analysis of GW231123 using a hierarchical merger model prefers the progenitors of the primary (secondary) 
    to have masses $137^{+20}_{-19} (103^{+20}_{-52}) \msun$
and spin magnitudes $0.9^{+0.1}_{-0.19}$ ($0.8^{+0.2}_{-0.51})$
    \cite{GW231123}.
Though we did not include spin angles in our analysis due to the disagreement between
    the posteriors for different waveform models \cite{GW231123},
    we note the possibility of significant precessing spin in GW231123 is suggestive
    of a dynamical spheroid encounter \cite{mcfactsI}.
We conclude that the masses and spin magnitudes of GW231123 can be produced readily by repeated
    mergers in accretion disks of AGN, populated by BHs
    drawn from a heavily mass-segregated nuclear star cluster (NSC).

\cite{mcfactsI} found that higher generation BHs have spin magnitudes roughly consistent 
    with $\sim 0.7$ for 2g, $\sim 0.8$ for 3g, and $\sim 0.9$ for 4g and higher mergers.
For a heavily segregated BH IMF (and modest AGN lifetime) 
    we find that the masses and spins of the progenitors of 
    are most consistent with resulting from a merger between fourth and third generation (4g,3g)
    BHs.

If GW231123 originated in an AGN accretion disk, $\sim 6-7$ first generation BH likely
    contributed to the mass of the final remnant.
Considering that disk model \cite{mcfactsIII}, AGN lifetime \cite{mcfactsIII}, and BH IMF (E.g. Table \ref{tab:rates})
    can each affect the event rate by
    an order of magnitude, and GW231123 is a fairly typical high-generation merger (E.g. Figure \ref{fig:corner-plots}),
    we find no evidence for a ``missing'' population of low-mass high-spin BBH
    from the AGN channel.
The models presented in this work assume that all AGN have properties which scale directly with 
    SMBH / NSC mass, and have a scaled Sirko-Goodman \cite{SG03} disk profile simulated by \texttt{pAGN}
    \cite{pAGN}.
Future work may explore the impact of relaxing those or other assumptions (accretion, galaxy properties)
    on the population of predicted detections, in the context of LVK observations.

\begin{paragraph}{Data/Code Availability}
\mcfacts{} is available at \url{https://github.com/McFACTS/McFACTS}.
The specific versions of software including \mcfacts{} used in this work,
    as well as all data are available at
    \url{https://doi.org/10.5281/zenodo.16895926}.

\end{paragraph}

\begin{acknowledgments}
The authors would like to thank Katelyn Breivik for expert feedback on the
    \cosmic{} IMF description in Section \ref{sec:models}.
This material is based upon work supported by NSF's LIGO Laboratory which is a major facility fully funded by the National Science Foundation.

VD is supported by an appointment to the NASA Postdoctoral Program at the NASA Goddard Space Flight Center administered by Oak Ridge Associated Universities under contract NPP-GSFC-NOV21-0031.
BM, KESF and HEC are supported by NSF AST-2206096. BM \& KESF are supported by NSF AST-1831415 and Simons Foundation Grant 533845 as well as Simons Foundation sabbatical support to release \mcfacts{}. The Flatiron Institute is supported by the Simons Foundation.
KN acknowledges support from the LSSTC Data Science Fellowship Program, which is funded by LSSTC, NSF Cybertraining Grant no. 1829740, the Brinson Foundation, and the Moore Foundation.
ROS acknowledges support from NSF PHY 2012057, PHY 2309172, and AST 2206321.

We acknowledge software packages used in this publication,
    including
    \texttt{NumPy} \cite{harris2020array},
    \texttt{SciPy} \cite{2020SciPy-NMeth}, 
    \texttt{Matplotlib} \cite{Hunter_2007}, 
\texttt{Astropy} \citeAstropy{},
    and \texttt{H5py} \cite{collette_python_hdf5_2014}.
This research was done using resources provided by the 
    Open Science Grid \citeOSG,
    which is supported by the National
    Science Foundation awards \#2030508 and \#1836650,
    and the U.S. Department of Energy's Office of Science. 
\end{acknowledgments}

\bibliography{Bibliography.bib}
\bibliographystyle{abbrv}

\end{document}